\documentclass{article}

\usepackage{preamble}

\newcommand{\degree}{\relax\ifmmode^\circ\else$^\circ$\fi}

\newcommand{\BLADED}{\texttt{BLADED}}
\newcommand{\x}{\ensuremath\mathbf{x}}
\newcommand{\y}{\ensuremath\mathbf{y}}
\newcommand{\A}{\ensuremath\mathcal{A}}
\newcommand{\E}{\ensuremath\mathbb{E}}

\newcommand{\Iref}{\ensuremath{}I_\mathrm{ref}}
\newcommand{\Vref}{\ensuremath{}V_\mathrm{ref}}
\newcommand{\Vhub}{\ensuremath{}V_\mathrm{hub}}
\newcommand{\Vdir}{\ensuremath{}\theta_\mathrm{wind}}
\newcommand{\Hs}{\ensuremath{}H_\mathrm{s}}
\newcommand{\Tp}{\ensuremath{}T_\mathrm{p}}
\newcommand{\Wdir}{\ensuremath{}\theta_\mathrm{wave}}
\newcommand{\VWdir}{\ensuremath{}\theta_\mathrm{wind/wave}}

\title{Fatigue design load calculations of the offshore NREL~5MW benchmark turbine using quadrature rule techniques}

\author[1,2]{L.M.M.~van~den~Bos\footnote{Corresponding author: \texttt{l.m.m.van.den.bos@cwi.nl}}}
\author[2]{W.A.A.M.~Bierbooms}
\author[3]{A.~Alexandre}
\author[1]{B.~Sanderse}
\author[2]{G.J.W.~van~Bussel}

\affil[1]{Centrum~Wiskunde~\&~Informatica, P.O.~Box 94079, 1090~GB, Amsterdam, The Netherlands}
\affil[2]{Delft~University~of~Technology, P.O.~Box 5, 2600~AA, Delft, The Netherlands}
\affil[3]{DNV\,GL, One~Linear~Park, Avon~Street, Temple~Quay, Bristol, BS2~0PS, United~Kingdom}

\begin{document}
\maketitle

\begin{abstract}
\noindent A novel approach is proposed to reduce, compared to the conventional binning approach, the large number of aeroelastic code evaluations that are necessary to obtain equivalent loads acting on wind turbines. These loads describe the effect of long-term environmental variability on the fatigue loads of a horizontal-axis wind turbine. In particular Design Load Case 1.2, as standardized by IEC, is considered. The approach is based on numerical integration techniques and, more specifically, quadrature rules. The quadrature rule used in this work is a recently proposed ``implicit'' quadrature rule, which has the main advantage that it can be constructed directly using measurements of the environment. It is demonstrated that the proposed approach yields accurate estimations of the equivalent loads using a significantly reduced number of aeroelastic model evaluations (compared to binning). Moreover the error introduced by the seeds (introduced by averaging over random wind fields and sea states) is incorporated in the quadrature framework, yielding an even further reduction in the number of aeroelastic code evaluations. The reduction in computational time is demonstrated by assessing the fatigue loads on the NREL~5MW reference offshore wind turbine in conjunction with measurement data obtained at the North Sea, both for a simplified and a full load case.
\end{abstract}

\section{Introduction}
For the certification of the design of a wind turbine, various load cases have been formulated in the IEC~61400 standard for both onshore~\cite{iec61400-1} and offshore~\cite{iec61400-3} wind turbines. These load cases vary from regular power production in commonly observed environmental conditions, to extreme conditions simulating storm and failure. All cases have one property in common: the aeroelastic simulation code that models the wind turbine must be evaluated a large number of times, which is computationally costly. In this work arguably one of the most costly load cases is considered: Design Load Case (DLC) 1.2, describing fatigue loading for a power producing offshore wind turbine under regular environmental conditions. As the time span of this load case is the full lifetime of the turbine, a large number of model runs is necessary to assess the statistics of the fatigue loading.

The conventional approach to assess the effect of environmental variability on the turbine lifetime is to firstly split the domain of the variables describing the environmental conditions in bins, secondly run the aeroelastic model several times in each bin (the so-called \emph{seeds}), and finally determine the quantity of interest (e.g.\ the weighted equivalent load) incorporating the probability of occurrence of each bin. This approach is suggested in the aforementioned standard~\cite{iec61400-1,iec61400-3} and has been successfully applied in previous research~\cite{Agarwal2008a,Freudenreich2008,Ragan2007}.

If applicable, binning is a versatile and effective tool to assess the effect of parameter variability. However, a major disadvantage of binning is that the number of necessary evaluations of the aeroelastic model is often prohibitively large, which limits its applicability. Several approaches have been suggested over the years to alleviate the high computational cost, e.g.\ lumping as discussed in~\citet[Section~8.3]{Kuehn2001}. This often requires non-trivial preprocessing steps and restrictive assumptions on the aeroelastic model.

Other approaches with similar flexibility exist. Examples include for example quasi Monte Carlo methods~\cite{Caflisch1998,Niederreiter1992}, surrogate methods~\cite{Babuska2007,Najm2009,Xiu2010}, or numerical integration techniques by means of quadrature approaches~\cite{Brass2011,Doostan2011,Gerstner1998,Ma2009a,Bos2016b}. These approaches have in common that regularity in the model (i.e.\ continuous responses to variability in the input parameters) is leveraged such that fast convergence can be obtained for sufficiently smooth functions. They have been successfully used to assess uncertainties in models in wind energy~\cite{Foti2016a,Graf2017,Murcia2018,Murcia2015,Schroeder2018,Bos2018a}, but a rigorous approach to model standardized load cases that directly competes with binning is still lacking. Furthermore, since the number of bins is fixed, it is difficult to assess the accuracy of the obtained estimation of the quantity of interest (the loads in the cases studied in this work). Moreover, explicit distributions of the parameters must be provided a priori. These are often unavailable, as there are various distributions that could potentially fit the measurements of the uncertain environment~\cite{Morgan2011}.

The goal of this work is to use quadrature rule approaches to assess fatigue loads described by DLC~1.2, leading to a competitive and attractive alternative approach to binning. Quadrature rules have the advantage that they are tailored to calculating integrals, which is the key goal of the load case. Moreover, they are versatile and robust by providing fast convergence for smooth functions, while still yielding accurate results for non-smooth functions. The quadrature rule used in this work is the implicit quadrature rule~\cite{Bos2018b}, which has the additional advantage over conventional quadrature rules~\cite{Doostan2011,Gerstner1998,Ma2009a} that no explicit knowledge about the distribution of the environmental parameters is required: only measurements of the environmental conditions are necessary. Furthermore, it provides an error estimate of the estimated load.

This article is structured as follows. In Section~\ref{sec:loadcase} fatigue load calculation is introduced, consisting of describing DLC~1.2, the on-site conditions considered in this article, and the wind turbine under consideration. The propagation of parametric uncertainty through an aeroelastic model is discussed in Section~\ref{sec:UP}. Here, the benchmark binning procedure is briefly explained and the quadrature rule algorithm and its most relevant mathematical properties are discussed in detail. The results obtained by binning and quadrature rules are compared in Section~\ref{sec:results}, demonstrating the applicability and prospect of quadrature rules. The article is concluded in Section~\ref{sec:conclusion}.

\section{Fatigue load calculation}
\label{sec:loadcase}
The DLCs as specified in the design requirements for offshore wind turbines by IEC~\cite{iec61400-1,iec61400-3} can be globally put into two categories: ultimate and fatigue load cases. The ultimate load cases describe the simulation of failure of a component due to rare events. Often these cases require nontrivial extrapolation procedures to assess the statistics. On the other hand, the fatigue load cases describe the simulation of regular environmental conditions over a long period of time to assess the effect of wear of the turbine. The last group is the focus of this article, as this type of cases requires a large number of evaluations of an aeroelastic code. In particular, the focus is on DLC~1.2 in combination with DLC~6.4. DLC~1.2 accounts for environmental conditions that result into power production and DLC~6.4 describes similar environmental conditions, with the sole exception that the turbine is idling due to too small or too large wind speeds\footnote{Strictly speaking, DLC~6.4 only accounts for wind speeds that are significantly smaller than the reference wind speed of the turbine ($\Vhub < 0.7\, \Vref$). We will also consider wind speeds larger than the cut-out speed, which also describe an idling wind turbine.}. There are various other fatigue load cases describing specific scenarios, e.g.\ start up (DLC~3.1), shutdown (DLC~4.1), and fault occurrence (DLC~2.4), but these are not the primary focus of this work since assessing these cases is significantly less computationally costly.

In this section, the specific details of the load case considered in this article are outlined. In Section~\ref{subsec:dlc12} the details of the DLC are discussed, as standardized by IEC~\cite{iec61400-1,iec61400-3}. In Section~\ref{subsec:mmij} and \ref{subsec:nrel5mw} the environmental conditions and the wind turbine under consideration are discussed. To keep the simulation as realistic and as reproducible as possible, freely available measurement data~\cite{Werkhoeven2015} and the NREL~5MW reference turbine~\cite{Jonkman2009} are used to describe the environment at the offshore site and model a wind turbine that should perform well at the location that is considered. In this work, the aeroelastic code \BLADED\ is employed for the load calculations, which is briefly discussed in Section~\ref{subsec:bladed}.

\subsection{Design Load Case 1.2: fatigue analysis}
\label{subsec:dlc12}
The goal of DLC~1.2 and DLC~6.4 is to assess the effect of wear on a wind turbine over a long period of time under normal design situations. All conditions are equal for both load cases, except for the wind condition that describes an idling turbine in DLC~6.4. Basically DLC~6.4 covers the cases where the wind speed at hub height averaged over 10 minutes, denoted as $\Vhub$, is smaller than the cut-in speed or larger than the cut-out speed and DLC~1.2 covers all other cases. It is convenient to bundle both cases in one single scenario and make no assumption about $\Vhub$.

Given the mean wind speed, the turbulence intensity is defined by means of the \emph{normal turbulence model} (NTM), which implies that the turbulence standard deviation $\sigma_1$ is given by the 90\% quantile for the given wind speed at hub height, i.e.\
\begin{equation}
	\sigma_1 = \Iref \, (0.75 \, \Vhub + b), \text{ with $b = \SI{5.6}{m/s}$}.
\end{equation}
The turbulence intensity, which should be used to generate an inflow wind field using the methods from Veers~\cite{Veers1988} or Mann~\cite{Mann1998}, then equals $\sigma_1 / \Vhub$. The lateral component of the turbulence standard deviation, denoted as $\sigma_2$, and the upward component, denoted as $\sigma_3$, are defined as $\sigma_2 = 0.8\,\sigma_1$ and $\sigma_3 = 0.5\,\sigma_1$ respectively (here, it is assumed that the Kaimal model is used). $\Iref$ denotes the expected value of the turbulence intensity at \SI{15}{m/s}, with $\Iref = 0.16$ for the case described in this article.

The wind direction $\Vdir$ with respect to the turbine is assumed to be uniformly distributed between \SI{-12}{\degree} and \SI{12}{\degree}, as the fatigue load case under consideration does not model extreme yaw misalignment. Notice that wind direction \SI{0}{\degree} is equivalent to no yaw misalignment. The wind direction and wind speed are independently distributed random variables.

The sea state is incorporated using a joint probability distribution of the significant wave height $\Hs$, the peak spectral period $\Tp$, and the wind at hub height $\Vhub$. A normal sea state is considered, so no anomalous events are incorporated and currents are ignored. The wind and wave directionality (and their dependence) have to be incorporated as co-directional or multi-directional (i.e.\ they are either equal or not). In this work, the latter option is chosen. For this purpose, the wave direction $\Wdir$ (with respect to the turbine) is used to determine the wind/wave misalignment:
\begin{equation}
	\label{eq:misalignment}
	\VWdir \coloneqq \Vdir-\Wdir.
\end{equation}
The distribution of the wind/wave misalignment can be determined by subtracting the stochastic wave direction (which should be inferred from the site conditions) from the uniformly distributed wind direction.

Combining the parameters describing the wind and sea conditions yields a five dimensional parameter space, called $\Omega$ in this work, consisting of all possible combinations of those parameters, called $\mathbf{x} \in \Omega$ in this work. It consists of the wind speed at hub height denoted by $\Vhub$, the wind direction denoted by $\Vdir$, the significant wave height denoted by $\Hs$, the peak spectral period denoted by $\Tp$, and the wind/wave misalignment denoted by $\VWdir$. Hence $\mathbf{x} = (\Vhub, \Vdir, \Hs, \Tp, \VWdir)^\mathrm{T}$. The concrete goal of DLC~1.2 and DLC~6.4 is to assess the effect of regular, long-term variability of these parameters on the forces acting on the wind turbine.

\subsection{Meteorological mast IJmuiden measurements}
\label{subsec:mmij}
\begin{figure}[t]
	\centering
	\begin{minipage}[t]{.5\textwidth}
		\centering
		\includegraphics[height=.8\textwidth]{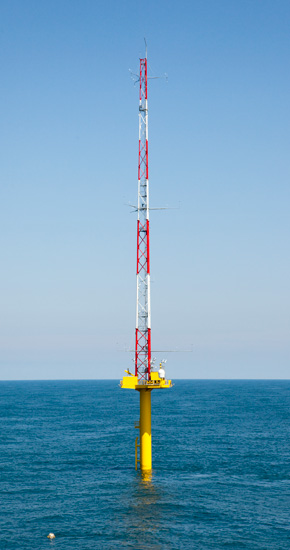}
		\caption{The meteorological mast IJmuiden}\label{fig:mmij}
	\end{minipage}%
	\begin{minipage}[t]{.5\textwidth}
		\centering
		{\footnotesize\includepgf{1.3\textwidth}{0pt}{turbine.tikz}}
		\caption{NREL~5MW reference wind turbine}\label{fig:nrel5mw}
	\end{minipage}
\end{figure}
The environmental conditions used for the load case calculations are based on the measurements done using meteorological mast IJmuiden~\cite{Werkhoeven2015}. This meteorological mast is situated approximately 85 km from the Dutch coast (at coordinates $52\degree 50' 53'' \mathrm{N} ~ 3\degree 26' 8'' \mathrm{E}$). At the measurement site the water depth is approximately 28 meters, see Figure~\ref{fig:mmij} for an illustration. The measurement campaign started in November 2011 and lasted until March 2016. Up to some interruptions, the wind conditions at various heights, wave conditions, and ocean current conditions at various depths have been obtained using a lidar and a buoy throughout that period of time.

For the purpose of this work, we are mainly interested in the statistical behavior of the wind in relation with the sea state. To this end, a straightforward preprocessing step is performed to combine the hourly measured sea state with the 10-minute averages of the wind conditions at that specific moment. After removal of all invalid measurements (due to measurement errors or interruptions), 24\,650 samples of the wind speed and wind direction at hub height (which is \SI{90}{m}, see the next section) and significant wave height, wave direction, and peak spectral period are obtained. These samples of the combined wind and wave conditions constitute 88\% of all available measured wave directions (measured hourly) and 10\% of all available measured wind directions (measured every 10 minutes).

\begin{figure}[t]
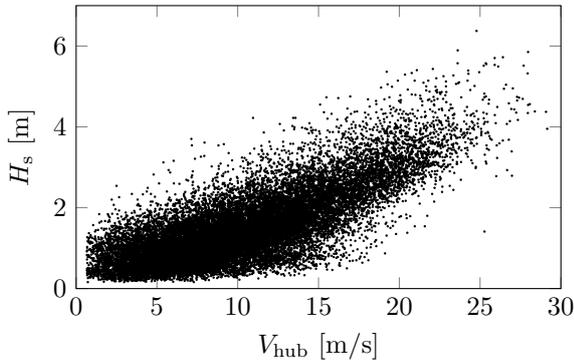
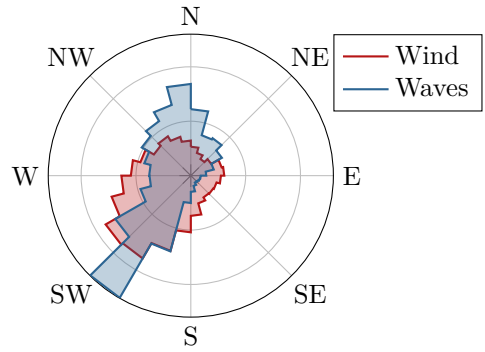

	\centering
	\begin{minipage}{.5\textwidth}
		\centering
		\includepgf{\textwidth}{.67\textwidth}{vhub-Hs.tikz}
		\subcaption{Mean wind speed and significant wave height}
		\label{subfig:measurementsa}
	\end{minipage}%
	\begin{minipage}{.5\textwidth}
		\centering
		\includepgf{.67\textwidth}{.67\textwidth}{wdir-vdir.tikz}
		\subcaption{Wind and wave direction}
		\label{subfig:measurementsb}
	\end{minipage}
	\caption{Measurements of the meteorological mast IJmuiden. \emph{Left:} all measurements of the wind speed and significant wave height, demonstrating their mutual dependence. \emph{Right}: Wind rose illustrating a histogram of the wind and wave direction. Both have a dominant south west inflow.}
	\label{fig:measurements}
\end{figure}

The obtained measurements are visualized in Figure~\ref{fig:measurements}. Notice that Figure~\ref{subfig:measurementsa} clearly shows that there are no measurements with $\Vhub = \SI{0}{m/s}$, i.e.\ the measurement devices have a small tolerance resulting in a measured wind speed that is unconditionally positive. Figure~\ref{subfig:measurementsb} demonstrates that the dominant wind and wave direction is south west. There is also a significant number of waves from north, possibly due to the open connection with the North Sea at the offshore site of the meteorological mast. Nonetheless, waves from north rarely occur in combination with wind from south west, as the wind/wave misalignment has approximately zero mean.

\subsection{NREL~5MW reference offshore wind turbine}
\label{subsec:nrel5mw}
The NREL~5MW reference wind turbine for offshore system development is a fictional offshore wind turbine designed to support concept studies aimed at assessing offshore wind technology~\cite{Jonkman2009}. It is a horizontal-axis wind turbine rated at \SI{5}{MW} with a rotor diameter of \SI{126}{m}. Its hub height is \SI{90}{m}, which allows for the straightforward usage of the measurement data discussed in the previous section, as measurements at \SI{90}{m} are available. The cut-in, rated, and cut-out wind speed of the turbine equal \SI{3}{m/s}, \SI{11.4}{m/s}, and \SI{25}{m/s} respectively. As the mean wind speed of the meteorological mast at \SI{90}{m} is \SI{10.13}{m/s}, simulation of the NREL~5MW wind turbine at the location of the IJmuiden mast describes a realistic scenario of a power producing wind turbine. The geometry of the turbine is depicted in Figure~\ref{fig:nrel5mw}.

\subsection{Aeroelastic code BLADED}
\label{subsec:bladed}
The simulation of the turbine is performed using the aeroelastic wind turbine modeling tool \BLADED\ (version 4.6). One run of \BLADED\ to obtain the forces acting on the turbine over a period of 10 minutes takes approximately 50 minutes on the hardware used for the numerical experiments, which significantly dominates the total computation time. All other algorithms discussed in this article (such as determining the bins, the quadrature rules, and all post processing procedures) are negligible in this regard.

Given specific values of the environmental parameters, \BLADED\ is used firstly for simulating the wind turbine to obtain the forces and secondly for post processing the obtained results. Obtaining forces consists of two steps. Firstly, a turbulent wind field using the average wind speed at hub height and the turbulence intensity is generated. Then secondly, using this turbulent wind field, the random sea state, and the design of the NREL~5MW wind turbine, a time history of the forces acting on various components of the turbine is calculated. By repeatedly doing such calculations for various environmental conditions, and incorporating the frequency of occurrence of each specific set of environmental parameter values, weighted equivalent loads are calculated for various representative slopes of the S--N curve. An equivalent load is a representative load of a specific load time history resulting in the same damage. A weighted equivalent load is an equivalent load where the frequency of occurrence of several time histories is taken into account. Determining these loads is the goal of this work, which is a standardized way of reporting and assessing wind turbine performance.

An equivalent load is calculated using the time history of the forces acting on the turbine. It is out of the scope of this article to fully discuss all details, we refer interested readers to~\citet{Burton2008} and the references therein for a general introduction or the Theory~Manual of \BLADED~\cite{DNV2014b} for the specific implementation details within \BLADED.

There exists variability in the output of \BLADED\ due to the random nature of the turbulent wind field and sea state. To model these, it is common to generate several wind fields and sea states and average the time histories of the forces over these realizations. These so-called \emph{seeds} are also incorporated in the framework in this article.

It is important to emphasize that the procedure of determining equivalent loads and calculating the seeds is similar for both binning, which is the well-known conventional approach, and numerical integration by means of quadrature methods, which is the novel approach requiring significantly less \BLADED\ evaluations suggested in this article. No modifications to the code or input files of \BLADED\ are necessary, except for the frequency of occurrence of each individual simulation. This is a major advantage of the approach suggested in this article, as all existing well-developed frameworks can be used without significant changes (i.e.\ the approach is \emph{non-intrusive}).

\section{Numerical integration for load calculations}
\label{sec:UP}
The procedure to determine all equivalent loads that are necessary to obtain weighted equivalent loads is the most costly part of DLC~1.2, as it requires the calculation of a large number of time histories of loads. In this section, it is demonstrated why this happens in case binning is used and a numerical integration framework is proposed as efficient and accurate alternative.

To introduce equivalent loads in our framework, let $u: \Omega \to \mathbb{R}$ be the map that yields an equivalent load (the quantity of interest) for specific values of the environmental conditions $\mathbf{x} \in \Omega$ (see Section~\ref{subsec:dlc12}). To incorporate the uncertain nature of the environment, the equivalent loads are combined into a single weighted equivalent load, which is ideally calculated as follows:
\begin{equation}
	\label{eq:DELint}
	L^\mathrm{eq}_{\phantom{N}} = \sqrt[m]{\E[u(\x)^m]}, \text{ with } \E[u(\x)] = \int_\Omega u(\x) \, \rho(\x) \dd \x.
\end{equation}
Here, $\rho(\x)$ is the probability density function modeling the uncertainty of the environment. The integer $m$ is the slope of the S--N curve, where it is implicitly assumed that the S--N curve can be represented by a monomial with power $m$. The notation $L^\mathrm{eq}_{\phantom{N}}$ is used to represent this (ideal) weighted equivalent load.

In the case studied in this article, it is not assumed that there is a known probability density function of the environmental parameters, but instead measurements are directly used (see Section~\ref{subsec:mmij}). In this case, $L^\mathrm{eq}_{\phantom{N}}$ can be introduced as follows:
\begin{equation}
	\label{eq:DELsum}
	L^\mathrm{eq}_{\phantom{N}} = \sqrt[m]{\E[u(\x)^m]}, \text{ with } \E[u(\x)] = \frac{1}{K} \sum_{k=0}^{K-1} u(\y_k),
\end{equation}
where $\y_k \in \Omega$ are all measurement points available (as depicted in Figure~\ref{fig:measurements}) and $K = 24\,650$ depicts the number of measurement points. Each $\y_k$ is a 5-dimensional vector describing the measured parameters (see Section~\ref{subsec:mmij}). Both \eqref{eq:DELint} and \eqref{eq:DELsum} can be used in the framework proposed here, i.e.\ either measurement data or a distribution can be used straightforwardly.

Usually it is not viable to evaluate \eqref{eq:DELint} exactly, as $u$ is a map that involves evaluating a complex aeroelastic model. Also \eqref{eq:DELsum} cannot be used, as it is intractable to evaluate the aeroelastic model for each measurement point. Therefore, the integral is replaced by a weighted sum, obtaining the following approximation:
\begin{equation}
	\label{eq:DELquad}
	L^\mathrm{eq}_N = \sqrt[m]{\A_N[u(\x)^m]}, \text{ with } \A_N[u(\x)] = \sum_{k=0}^N u(\x_k) w_k.
\end{equation}
where $\x_0, \dots, \x_N$ are $N+1$ specific representative values of the environmental parameters, called \emph{nodes} in this article, and $w_0, \dots, w_N$ the \emph{weights}. The weights model the frequency of occurrence of the nodes. The notation $L^\mathrm{eq}_N$ is used to represent a weighted equivalent load using $N$ runs of \BLADED. Ideally, we would have that $L^\mathrm{eq}_N \to L^\mathrm{eq}_{\phantom{N}}$ for $N \to \infty$ (if $L^\mathrm{eq}_{\phantom{N}}$ is according to \eqref{eq:DELint}) or $N \to K$ (if $L^\mathrm{eq}_{\phantom{N}}$ is according to \eqref{eq:DELsum}).

The standardized approach to determine the nodes and weights is by means of binning. In that case, $\x_k$ are equidistantly spaced and $w_k$ equal the frequency of occurrence of each environmental condition. This approach is discussed briefly in Section~\ref{subsec:binning}.

One major disadvantage of binning is the large number of nodes (and concomitant model runs) that are obtained for the five-dimensional parameter space considered in DLC~1.2. An alternative is to determine the nodes and weights such that they form an interpolatory quadrature rule. This yields an accurate estimation with relatively small number of nodes. The challenge is to construct the rule that incorporates the measurement data without reconstructing a (possibly inaccurate) probability density function and without requiring model evaluations at all samples $\y_k$. The method used in this work is discussed Section~\ref{subsec:quadrules}.

In \eqref{eq:DELint}, \eqref{eq:DELsum}, and \eqref{eq:DELquad} it is assumed that the function $u$ can be evaluated exactly. This is generally not the case, as the obtained equivalent loads still depend on the generated wind field and random sea. This is often solved by using \emph{seeds}, which consists of constructing various wind and sea states with similar environmental parameters and averaging the obtained equivalent loads. Since the obtained equivalent loads are used in a weighted sum, it is a natural idea to use more seeds for nodes with high weight and vice versa. With this idea, seeds can be incorporated in the quadrature rule framework. Details are discussed in Section~\ref{subsec:seedbalancing}.

\subsection{Binning}
\label{subsec:binning}
Binning is the conventional approach to assess variability in the wind and waves. The idea is to evaluate the aeroelastic code at equidistantly placed locations and post process the results using the probability density function of the variables under consideration. The bins can be interpreted as quadrature rule nodes, yielding a rule with equidistant $\x_k$. The weights $w_k$ describe the frequency of occurrence of $\x_k$.

If a distribution $\rho(\x)$ is known, the weight of the $k$-th bin $b_k$ is typically determined as follows~\cite[Annex~G]{iec61400-1}:
\begin{equation*}
	w_k = \int_{b_k} \rho(\x) \dd \x.
\end{equation*}
If measurements are considered, the weights are the fraction of measurements in the bin of $\x_k$, i.e.
\begin{equation*}
	w_k = \frac{K_k}{K},
\end{equation*}
where $K_k$ is the number of measurements in the bin of $\x_k$ and $K$ is the total number of measurements.

\begin{figure}
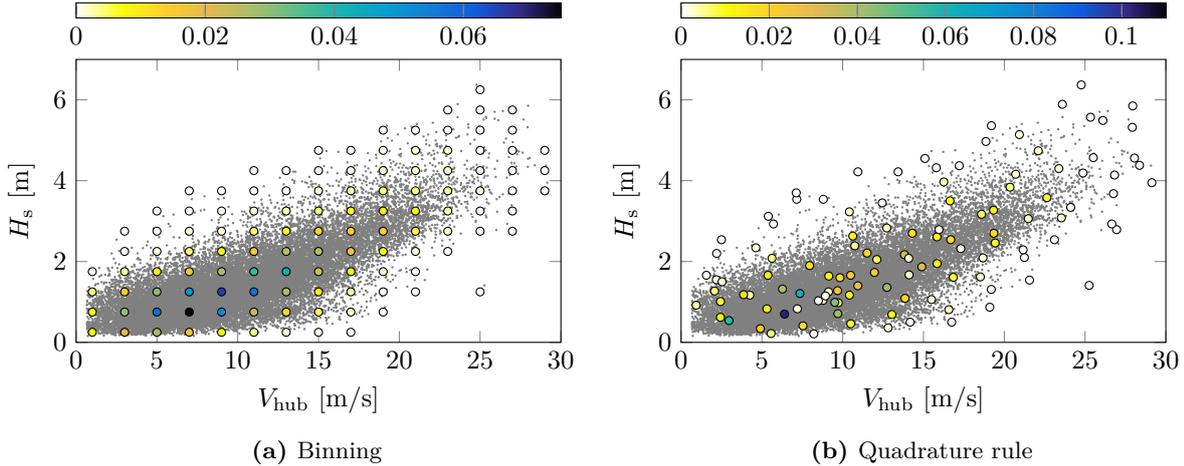

	\begin{minipage}{.5\textwidth}
		\centering
		\includepgf{\textwidth}{.67\textwidth}{binning.tikz}
		\subcaption{Binning}
		\label{subfig:binning}
	\end{minipage}%
	\begin{minipage}{.5\textwidth}
		\centering
		\includepgf{\textwidth}{.67\textwidth}{implquad.tikz}
		\subcaption{Quadrature rule}
		\label{subfig:implquad}
	\end{minipage}
	\caption{The representative values of the wind speed at hub height and significant wave height if binned or if determined by means of a quadrature rule. The color of the nodes indicates their weight.}
	\label{fig:quadrules}
\end{figure}

An example of a quadrature rule obtained by applying binning is illustrated in Figure~\ref{subfig:binning}, where the same measurement points as in Figure~\ref{subfig:measurementsa} have been used to determine the bins. The weights are illustrated by means of colors and nodes with zero weight are not depicted. The mean wind speed at hub height is binned with bin sizes of \SI{2}{m/s} and the significant wave height is binned with bin sizes of \SI{0.5}{m}. Notice that many weights equal zero, requiring no concomitant model runs. Nonetheless, the majority of the nodes (117 out of 210) has possibly small, but positive weight.

The advantage of this approach is that it is versatile and straightforward to implement. Moreover it is supported by many software packages for wind turbine simulation. However, if multiple environmental parameters are considered the number of bins increases exponentially. To see this, notice that if $d$ parameters are considered with $B$ bins in each direction, the total number of bins is proportional to $B^d$, even if bins with zero weight are neglected. This severely limits the applicability of binning if a large number of parameters is considered.

\subsection{Interpolatory quadrature rules}
\label{subsec:quadrules}
As an alternative to binning, we consider interpolatory quadrature rules. The nomenclature and mathematical framework of these rules is introduced in Section~\ref{subsubsec:quadrules}. In Section~\ref{subsubsec:implquad} the quadrature rule used in this work is explained.

\subsubsection{Nomenclature}
\label{subsubsec:quadrules}
The key property of an interpolatory quadrature rule is that it is constructed such that it integrates all polynomials up to a certain degree exactly. Hence if $\x_0, \dots, \x_N$ and $w_0, \dots, w_N$ are the nodes and weights of a quadrature rule of degree $D$, then
\begin{equation*}
	\A_N[\varphi] = \sum_{k=0}^N \varphi(\x_k) w_k = \int_\Omega \varphi(\x) \, \rho(\x) \dd \x = \E[\varphi],
\end{equation*}
for all polynomials $\varphi$ of degree $D$ or less. The degree of a multivariate polynomial is determined as the \emph{sum} of its powers, e.g.\ $\varphi(\x) = \Vhub^3 \, \Vdir^7$ has degree $10$.

The probability density function $\rho$ does not appear in the definition of the quadrature rule directly, but is incorporated in the nodes and the weights. If nodes $\x_0, \dots, \x_N$ are given, an interpolatory quadrature rule of degree $N$ can be obtained by solving the following linear system for the weights $w_0, \dots, w_N$:
\begin{equation}
	\label{eq:vandermonde}
	\begin{pmatrix}
		\varphi_0(\x_0) & \cdots & \varphi_0(\x_N) \\
		\vdots & \ddots & \vdots \\
		\varphi_N(\x_0) & \cdots & \varphi_N(\x_N)
	\end{pmatrix}
	\begin{pmatrix}
		w_0 \\
		\vdots \\
		w_N
	\end{pmatrix}
	=
	\begin{pmatrix}
		\mu_0 \\
		\vdots \\
		\mu_N
	\end{pmatrix}.
\end{equation}
Here, $\varphi_j$ are monomials, i.e.\ polynomials of only one term (e.g.\ the $\Vhub^3\,\Vdir^7$ example above). In the univariate case ($d=1$), it holds that $\varphi_j(\x) = \x^j$. Multivariate polynomials are sorted graded lexicographically~\cite{Xiu2010} to accommodate quadrature rules of any number of nodes $N$ (we will call such a rule, with a little abuse of nomenclature, a rule of degree $N$). The values $\mu_j$ are the raw moments of the distribution, defined as follows:
\begin{equation}
	\mu_j = \mathbb{E}[\varphi_j] = \int_\Omega \varphi_j(\x) \, \rho(\x) \dd \x, \text{ for $j = 0, \dots, N$},
\end{equation}
which uses \eqref{eq:DELint}. If measurements are used, the raw moments are defined as follows:
\begin{equation}
	\label{eq:moments}
	\mu_j = \frac{1}{K} \sum_{k=0}^{K-1} \varphi_j(\y_k), \text{ for $j = 0, \dots, N$},
\end{equation}
where $\y_k$ constitute all available measurements. The latter definition, which is employed in this work, uses \eqref{eq:DELsum}.

A quadrature rule of high degree is not necessarily accurate if applied to functions that are not a polynomial. On the other hand, if the linear system from \eqref{eq:vandermonde} is satisfied (i.e.\ the quadrature rule has degree $N$) \emph{and} all weights are non-negative (i.e.\ $w_k \geq 0$ for all $k$), then the following inequality holds~\cite{Brass2011}:
\begin{equation}
	\label{eq:lebesgue}
	| \E[u] - \A_N[u] | \leq 2 \quad \min_{\mathclap{\varphi \in \left< \varphi_0, \dots, \varphi_N \right>}} \quad \| \varphi - u \|_\infty, \text{ with } \| \varphi - u \|_\infty = \max_{\x \in \Omega} | \varphi(\x) - u(\x) |.
\end{equation}
This inequality is commonly known as the Lebesgue inequality. The right hand side of the inequality, i.e.\ $\min \| \varphi - u \|_\infty$, is the minimal distance between the aeroelastic model $u$ and \emph{any} polynomial $\varphi$. It is notoriously difficult to determine this distance~\cite{Watson1980} (even if the model is known analytically). This is not a major issue, since the bound tells that the error made by integration using an interpolatory quadrature rule is at most twice as bad as the smallest distance between the model and any polynomial, even if that distance is unknown.

In other words, if the computational model \emph{can} be approximated by means of a polynomial, a quadrature rule is a viable numerical integration methodology. A model can be approximated by a means of a polynomial if it is continuous and if its domain is bounded, which is the case for \BLADED\ and the parameter space $\Omega$ considered in this article. Many more convergence results considering quadrature rules and polynomial approximation exist. The interested reader is referred to~\citet{Brass2011} and~\citet{Watson1980} and the references therein for more information.

There exist various approaches to construct interpolatory quadrature rules with positive weights. For univariate cases, the Gaussian quadrature rules~\cite{Golub1969} and Clenshaw--Curtis quadrature rules~\cite{Clenshaw1960} are arguably the best-known rules. In higher dimensional spaces, constructing quadrature rules has similar limitations as when using binning: the number of nodes grows exponentially as the number of parameters grows. To alleviate this, the Smolyak sparse grid can be used~\cite{Novak1999,Smolyak1963}, although this approach requires that all parameters are independently distributed (which is clearly not the case, as described in Section~\ref{subsec:dlc12}).

In this work we use the implicit quadrature rule to determine the nodes and the weights~\cite{Bos2018b}. This quadrature rule is constructed as a subset of a provided sample set, so it is very suitable for the measurement data that is considered in this work. It has the advantage that its number of nodes does not grow exponentially fast if the number of parameters is increased, which is the case with binning. Moreover, it has positive weights and therefore yields an accurate estimation of $\E[u]$, whose error can be efficiently estimated. It is briefly explained in Section~\ref{subsubsec:implquad}.

An example of the implicit quadrature rule, determined using the measurement points from Figure~\ref{subfig:measurementsa}, is depicted in Figure~\ref{subfig:implquad}. The rule is chosen to consist of 117 nodes (the number of bins with non-zero weight). It is clearly visible that the nodes have positive weights and to a certain extent represent the distribution of the measurement points.

\subsubsection{Implicit quadrature rule}
\label{subsubsec:implquad}
Consider $K$ sample points $\y_0, \dots, \y_{K-1}$ to be given. The goal is to construct, for a desired number of nodes $N$ (with $N \ll K$), a quadrature rule with nodes $\x_0, \dots, \x_N$ (with $\x_k \in \{ \y_k \}$) and \emph{positive} weights $w_0, \dots, w_N$ such that
\begin{equation}
	\label{eq:keyproperty}
	\sum_{k=0}^N \varphi_j(\x_k) w_k = \frac{1}{K} \sum_{k=0}^{K-1} \varphi_j(\y_k), \text{ for $j = 0, \dots, N$}.
\end{equation}
Notice that the right hand side of this equality is formed by the raw moments from \eqref{eq:moments}, which are simply the averages of all $\varphi_j$ determined over the samples. As $\varphi_j$ are polynomials, this estimate can be determined for very large values of $K$ without any significant computational cost.

The key idea of the implicit quadrature rule is to interpret the right hand side of \eqref{eq:keyproperty} as a quadrature rule with $K$ nodes $\x_k = \y_k$ and weights $w_k = 1/K$ for all $k$. Then the Vandermonde-matrix of \eqref{eq:vandermonde} is not a square matrix, but is the following $(N+1) \times K$-matrix:
\begin{equation}
	\label{eq:hatV}
	V =
	\begin{pmatrix}
		\varphi_0(\x_0) & \dots & \varphi_0(\x_N) & \varphi_0(\x_{N+1}) & \dots & \varphi_0(\x_{K-1}) \\
		\vdots & \ddots & \vdots & \vdots & \ddots & \vdots \\
		\varphi_N(\x_0) & \dots & \varphi_N(\x_N) & \varphi_N(\x_{N+1}) & \dots & \varphi_N(\x_{K-1})
	\end{pmatrix}.
\end{equation}
This matrix has a larger number of columns than rows, so it has $(K-N-1)$ null vectors, i.e.\ a vector $\mathbf{c} = \trans{(c_0, \dots, c_{K-1})}$ such that $V \mathbf{c} = \mathbf{0} = \trans{(0, \dots, 0)}$. Such a null vector can be exploited to construct a smaller quadrature rule~\cite{Bos2016b}, by noticing that
\begin{equation*}
	\sum_{k=0}^N u(\x_k) (w_k - \alpha c_k) = \frac{1}{K} \sum_{k=0}^{K-1} u(\y_k),
\end{equation*}
where $\alpha \in \mathbb{R}$ is an arbitrary constant (any multiple of a null vector is a null vector). A smaller quadrature rule can be obtained by selecting $\alpha$ such that $w_k - \alpha c_k = 0$ for one $k$. The goal is to keep all weights positive, i.e.\ $w_k - \alpha c_k \geq 0$, which is achieved by choosing $\alpha$ as follows:
\begin{equation}
	\label{eq:alpha}
	\alpha = \min\left(\frac{w_k}{c_k} \mid c_k > 0\right).
\end{equation}
This value is firstly such that $w_k - \alpha c_k \geq 0$ for all $k$, secondly such that there exists a $k = k_0$ with $w_{k_0} - \alpha c_{k_0} = 0$, and finally such that $\{ w_k - \alpha c_k \}$ can be used as weights of the quadrature rule, since $\mathbf{c}$ is a null vector. Therefore, the $k_0$-th node and weight can be removed from the quadrature rule by updating the weights using $w_k \gets w_k - \alpha c_k$ (as $w_{k_0} - \alpha c_{k_0} = 0$).

Repeatedly applying this procedure to the quadrature rule with $K$ nodes leads eventually to a quadrature rule of degree $N$ with $N+1$ nodes, which was the original goal. Note that there are always two possible nodes that can be removed, as both $\mathbf{c}$ and $-\mathbf{c}$ are null vectors and \eqref{eq:alpha} only uses the positive values of the null vector. Eventually the rule consists of $N+1$ nodes and the matrix from \eqref{eq:hatV} becomes a square matrix. In that case no more nodes can be removed from the rule without also removing a function $\varphi_j$ that will be integrated accurately by the rule. This yields the nodes $\x_k$ and weights $w_k$ that can be used in \eqref{eq:DELquad} to accurately estimate equivalent loads.

The procedure to remove nodes can also be used to approximate the error made by a quadrature rule. The trick is to, given a quadrature rule, iteratively remove a function $\varphi_j$ (the last row of \eqref{eq:hatV}, this decreases the degree of the rule) and subsequently remove a node from the quadrature rule. Then a \emph{sequence} of quadrature rules with positive weights of decreasing degree and decreasing number of nodes is obtained, such that the following expression can be used to estimate the accuracy of the rules:
\begin{equation*}
	e_n = | \A_n[u] - \A_N[u] |.
\end{equation*}
Here, $n$ denotes the number of nodes of the rule upon removal of nodes (namely $N-n$ nodes). Determining $e_n$ for several increasing $n$ (with $n < N$) gives insight in the decay of the error made by the quadrature rule, as $e_n$ is bounded by the errors made using $\A_n$ and $\A_N$:
\begin{equation}
	\Bigl| |\A_n[u] - \E[u]| - |\A_N[u] - \E[u]| \Bigr| \leq e_n \leq \Bigl| | \A_n[u] - \E[u] | + | \A_N[u] - \E[u] | \Bigr|.
\end{equation}
The left hand side of the inequality describes that $e_n$ is larger than the difference of the errors made by $\A_n$ and $\A_N$. The right hand side of the inequality describes that $e_n$ is smaller than the sum of both errors. If the quadrature rule is applied to a function $u$ that can be numerically integrated using a quadrature rule (i.e.\ \eqref{eq:lebesgue} is a good error estimate), a small $e_n$ implies that the error of the quadrature rule is small. An even better estimation of the error can be obtained by repeatedly determining $e_n$ for different sequences of quadrature rules, and taking the average of the obtained values. This is possible as multiple sequences exist given the freedom in choosing $\mathbf{c}$ or $-\mathbf{c}$.

\subsection{Seed balancing}
\label{subsec:seedbalancing}
Both binning and quadrature rule approaches yield nodes that describe values of the environmental parameters that should be evaluated using the aeroelastic code. These evaluations have been denoted by the function $u(\x)$. However, these evaluations still contain a degree of randomness: for each set of parameters, a certain number of \emph{seeds} must be evaluated. This involves three steps: firstly various turbulent wind fields and sea states are constructed, secondly the loads are determined for \emph{each} wind field and sea state, and finally the obtained loads are averaged. Hence, if $S_k$ seeds are used to determine $u(\x_k)$, this can be denoted as follows:
\begin{equation*}
	u(\x_k) = u_{S_k}(\x_k) \coloneqq \frac{1}{S_k} \sum_{s=1}^{S_k} u^{(s)}(\x_k),
\end{equation*}
where $u^{(s)}(\x_k)$ is an evaluation of the aeroelastic code using the $s$-th wind field and $s$-th sea state. Ideally, we would like to accurately approximate $\E[u_\infty]$, with
\begin{equation*}
	u_\infty(\x) = \lim_{S \to \infty} \frac{1}{S} \sum_{s=1}^S u^{(s)}(\x).
\end{equation*}
The approximation used in this work is $\A_N[u_S]$ for finite $N$ and $S$. Here, $S = \sum_k S_k$ denotes the total number of \BLADED\ evaluations used in this quadrature rule estimate. It yields the following error estimate:
\begin{equation}
	\label{eq:totalerror}
	|\E[u_\infty] - \A_N[u_S]| \leq \underbrace{|\E[u_\infty] - \A_N[u_\infty]|}_{\text{Quadrature error}} + \underbrace{|\A_N[u_\infty] - \A_N[u_S]|}_{\text{Seed error}}.
\end{equation}
The focus in this section is on the seed error, since the quadrature error has been discussed extensively in Section~\ref{subsec:quadrules}. It is common to use a fixed number of seeds for each node (i.e.\ all $S_k$ are equal). However, an error of similar order of magnitude determined with a smaller total number of seeds can be obtained by selecting the number of seeds per node based on the frequency of occurrence of the node. In this way, the accuracy in the overall approximation of the integral is preserved with a reduced cost. In this section this intuition is derived mathematically.

First, notice that the seed error behaves as follows~\cite{Caflisch1998}:
\begin{equation}
	\label{eq:CLT}
	|\A_N[u_\infty] - \A_N[u_S]| \leq \sum_{k=0}^N |u_\infty(\x_k) - u_{S_k}(\x_k)| w_k, \text{ with } |u_\infty(\x_k) - u_{S_k}(\x_k)| \sim \frac{1}{\sqrt{S_k}}.
\end{equation}
Throughout this section we write $A \sim B$ if the growth or decay of $A$ is asymptotically the same as $B$. The constant of proportionality is the variance of the integrand, which typically depends on $\x_k$ and $m$ (the inverse S--N slope, recall that the integrand in \eqref{eq:DELquad} is $u^m$). It is omitted from this estimation, but can straightforwardly be incorporated in the procedure discussed in this section, provided that its dependence on $\x_k$ and $m$ is explicitly known.

The computational time of \BLADED\ does not depend on the specific wind field or sea state under consideration, so the total computational time scales linearly with the number of seeds. Hence given an accuracy goal $\mathcal{E}$, the goal is to determine $\vareps_{S_0}, \dots, \vareps_{S_N}$ that firstly minimize the total computational time $S$, denoted as the total number of seeds used for all calculations, and secondly achieve the accuracy goal, denoted by the seed error:
\begin{equation}
	\label{eq:minprob}
	\text{Minimize } S = \sum_{k=0}^N S_k \sim \sum_{k=0}^N \frac{1}{\vareps_{S_k}^2}, \text{ with } \vareps_{S_k} \coloneqq |u_\infty(\x_k) - u_{S_k}(\x_k)|, \text{ such that }
	\mathcal{E} \geq \sum_{k=0}^N \vareps_{S_k} w_k.
\end{equation}
Notice that for $N \to \infty$, the asymptotic error of this expression decays to 0. Even though the error estimate is asymptotic, the relation $S_k = \vareps_{S_k}^{-2}$ will be used throughout this work.

The problem sketched in \eqref{eq:minprob} is a constrained minimization problem, that can be solved using the method of Lagrange multipliers~\cite{Bertsekas1982}, which yields the solution
\begin{equation*}
	\vareps_{S_k} = A \left( \frac{w_k}{2} \right)^{-1/3},
\end{equation*}
where $A$ is a scaling constant such that $\sum_{k=0}^N \vareps_{S_k} w_k = \mathcal{E}$. Hence the number of seeds for the $k$-th node should ideally equal
\begin{equation}
	\label{eq:seeds}
	S_k = \frac{1}{\vareps_{S_k}^2} = A' w_k^{2/3}, \text{ with } A' = \frac{1}{2} \sqrt[3]{2} \, A^{-2} \approx 0.630 \, A^{-2}.
\end{equation}
The number of seeds $S_k$ is larger if the associated weight $w_k$ is larger and no seeds (hence no \BLADED\ runs) are necessary if $w_k$ is zero. The latter is straightforward to observe: if the weight of a node is zero, it does not contribute to the quadrature rule approximation. Notice that the relation between $S_k$ and $w_k$ is non-linear due to the non-linear relation between $\vareps_{S_k}$ and $S_k$.

Only in rare cases it holds that all values of $S_k$ as defined by \eqref{eq:seeds} are integers. It is difficult to solve the optimization problem stated here with the restriction that all $S_k$ are integer, so in this article the number of seeds is simply rounded up such that the accuracy goal is reached in all cases.

\begin{figure}
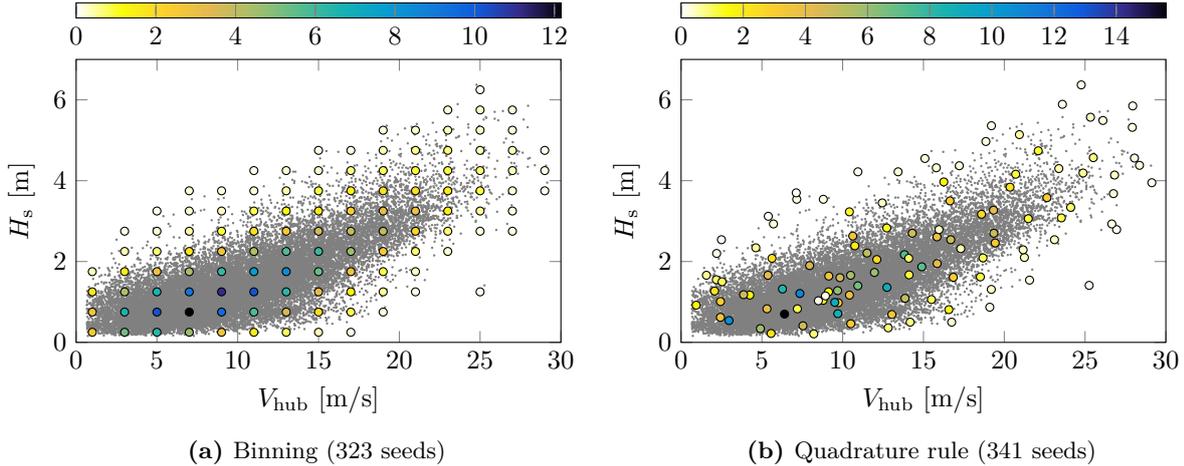

	\begin{minipage}{.5\textwidth}
		\centering
		\includepgf{\textwidth}{.67\textwidth}{seeds-binning.tikz}
		\subcaption{Binning (323 seeds)}
		\label{subfig:seeds-binning}
	\end{minipage}%
	\begin{minipage}{.5\textwidth}
		\centering
		\includepgf{\textwidth}{.67\textwidth}{seeds-implquad.tikz}
		\subcaption{Quadrature rule (341 seeds)}
		\label{subfig:seeds-implquad}
	\end{minipage}
	\caption{Number of seeds per node if the seeds are balanced with an accuracy goal $\mathcal{E} = 1/\sqrt{5}$. The color of the nodes indicates the number of seeds (before rounding up to the nearest integer).}
	\label{fig:seeds}
\end{figure}

\begin{table}
	\caption{Example of reduction in number of seeds when applying seed balancing. The number of nodes equals 117 for both binning and the implicit quadrature rule. The accuracy goal $\mathcal{E}$ equals $1/\sqrt{5}$.}\label{tbl:seedbalancing}

	\setlength\doublerulesep{2.0pt}

	\centering
	\begin{tabular}{r|ll}
		& \textbf{Unbalanced} & \textbf{Balanced} \\
		\hline
		\hline
		\textbf{Binning} & 585 & 323 (\SI{56}{\percent}) \\
		\textbf{Quadrature rule} & 585 & 341 (\SI{58}{\percent})
	\end{tabular}
\end{table}

The effect of the seeds and their dependence on the weights is illustrated in Figure~\ref{fig:seeds} and summarized in Table~\ref{tbl:seedbalancing}, where the quadrature rules from Figure~\ref{fig:quadrules} are reconsidered. Here it is assumed that by default five seeds are used per bin or node, which is equivalent to an accuracy goal of $\mathcal{E} = 1/\sqrt{5}$. If binning is used, the total number of seeds reduces from 585 (five per bin with non-zero weight) to 323, which is only 56\% of the original cost. If the proposed quadrature rule from Section~\ref{subsubsec:implquad} is used, the total number of seeds reduces to 341, which is 58\% of the original cost.

\section{Numerical results}
\label{sec:results}
In order to demonstrate the benefits of the proposed quadrature rule, the five-dimensional DLC 1.2 is considered and the approach is compared to binning. However, it is infeasible to assess the complete load case using binning, as it results in an excessive number of runs of \BLADED. Therefore, first two simplified test cases are considered for the comparison. The first test case, which is discussed in Section~\ref{subsec:explicit}, consists of assessing the accuracy of the quadrature rule with respect to binning if the ``model'' $u$ is a known test function. The second test case, which is discussed in Section~\ref{subsec:results2}, consists of the DLC~1.2 load case, where only the wind conditions are considered uncertain (and the sea state is fixed at nominal conditions). The full five-dimensional load case is then assessed purely by means of a quadrature rule and contains the five aforementioned parameters, modeling both the wind and the sea state. The results of this case (which follows the IEC~standard~\cite{iec61400-1,iec61400-3} to a large extent) are discussed in Section~\ref{subsec:results5}.
\subsection{Test functions}
\label{subsec:explicit}
\begin{table}
	\caption{The test functions from Genz~\citep{Genz1984}. All $d$-variate functions depend on the $d$-element vectors $\mathbf{a}$ and $\mathbf{b}$. The vector $\mathbf{b}$ is an offset parameter to shift the function. The vector $\mathbf{a}$ describes the degree to which the family attribute is present.}

	\setlength\doublerulesep{2.0pt}

	\centering
	\begin{tabular}{l l}
		\textbf{Integrand Family} & \textbf{Attribute} \\
		\hline
		\hline
		$u_1(x) = \cos\left(2\pi b_1 + \sum_{i=1}^d a_i x_i\right)$ & Oscillatory \\
		$u_2(x) = \prod_{i=1}^d \left(a_i^{-2} + (x_i - b_i)^2\right)^{-1}$ & Product Peak \\
		$u_3(x) = \left(1 + \sum_{i=1}^d a_i x_i\right)^{-(d+1)}$ & Corner Peak \\
		$u_4(x) = \exp\left(- \sum_{i=1}^d a_i^2 (x_i - b_i)^2 \right)$ & Gaussian \\
		$u_5(x) = \exp\left(- \sum_{i=1}^d a_i |x_i - b_i|\right)$ & $C_0$ function \\
		$u_6(x) = \begin{cases}
			0 &\text{if $x_1 > b_1$ or $x_2 > b_2$} \\
			\exp\left(\sum_{i=1}^d a_i x_i\right) &\text{otherwise}
		\end{cases}$ & Discontinuous
	\end{tabular}
	\label{tbl:genz}
\end{table}
The five-dimensional Genz test functions~\cite{Genz1984} are considered, which are six functions $u_1, \dots, u_6$ constructed specifically to assess the accuracy of integration methods (see Table~\ref{tbl:genz}). The goal is to calculate $\mathbb{E}[u]$ defined by measurement data (see \eqref{eq:DELsum}), i.e.\ the goal is to use binning and an implicit quadrature rule to estimate
\begin{equation*}
	\mathbb{E}[u_f] = \frac{1}{K} \sum_{k=0}^{K-1} u_f(\hat\y_k), \text{ for $f = 1, \dots, 6$}.
\end{equation*}
The data under consideration are the measurements from the IJmuiden meteorological mast (see Section~\ref{subsec:mmij}), scaled to the domain of definition of the Genz test functions. Therefore, $\hat\y_k$ denotes $\y_k$ after scaling of all data to $\Omega = {[0, 1]}^d$ (with $d = 5$). As the functions $u_f$ are known exactly, no seeds are employed in this test case.

\begin{figure}
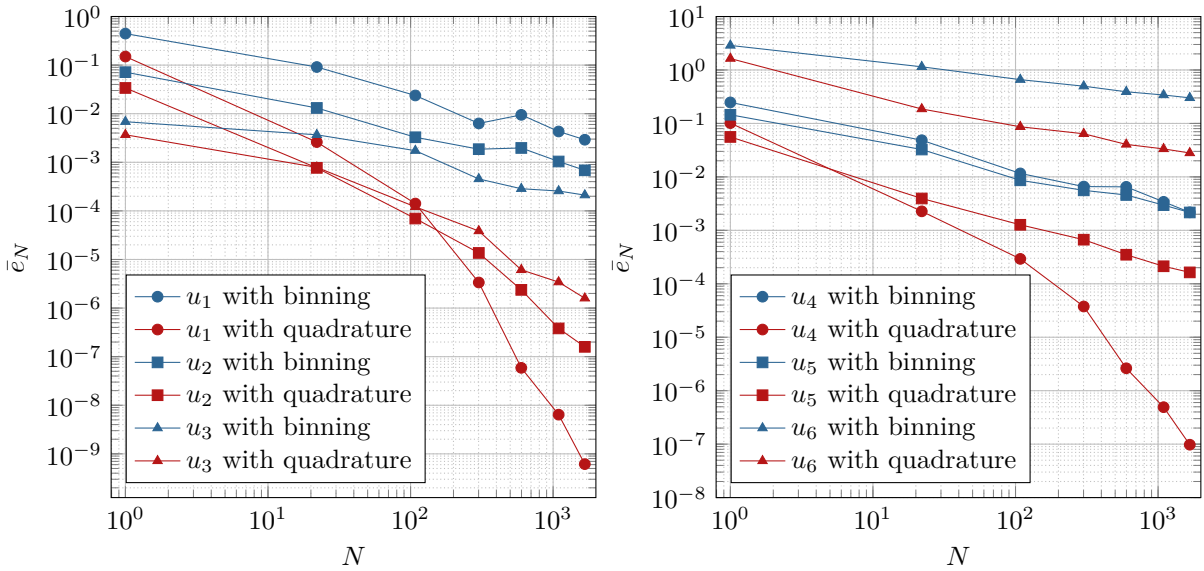

	\begin{minipage}[b]{.5\textwidth}
		\centering
		\includepgf{\textwidth}{\textwidth}{genz-123.tikz}
	\end{minipage}%
	\begin{minipage}[b]{.5\textwidth}
		\centering
		\includepgf{\textwidth}{\textwidth}{genz-456.tikz}
	\end{minipage}
	\caption{Mean integration error of the Genz test functions for various numbers of nodes, determined using binning (with the same number of bins in each dimension) or a quadrature rule.}
	\label{fig:genz}
\end{figure}

Binning (see Section~\ref{subsec:binning}) is performed by fixing a number $B = 1, \dots, 7$, constructing $B$ bins in each direction (obtaining $B^5$ bins), and removing all bins with zero weight. Then the implicit quadrature rule (see Section~\ref{subsubsec:implquad}) is constructed such that its number of nodes is equal to the number of obtained bins. The vectors $\mathbf{a}$ and $\mathbf{b}$, that define the Genz test function under consideration, are selected randomly in the unit hypercube and $\mathbf{a}$ is scaled such that $\| \mathbf{a} \|_2 = 5/2$. The experiment is repeated 100 times (i.e.\ with 100 random vectors $\mathbf{a}$ and $\mathbf{b}$) to obtain representative values of the integration error:
\begin{equation*}
	\bar{e}_N = \frac{1}{100} \sum_{k=1}^{100} e^{(k)}_N, \text{ with } e^{(k)}_N = | \mathbb{E}[u^{(k)}_f] - \mathcal{A}_N[u^{(k)}_f] |.
\end{equation*}
Here, $u^{(k)}_f$ is the $f$-th Genz test function constructed with the $k$-th pair of random vectors $\mathbf{a}$ and $\mathbf{b}$ and $N$ denotes the number of bins with non-zero weight or, equivalently, the number of nodes of the quadrature rule. The obtained results are depicted in Figure~\ref{fig:genz}.

First, notice that the quadrature rule consistently outperforms binning. The convergence behavior of binning is approximately equal for all functions, as binning does not leverage smoothness of the function due to its local approximation. Moreover a large number of bins is necessary to obtain an accurate estimation. On the other hand, the convergence behavior of the quadrature rule clearly reflects the attributes of the test functions (see Table~\ref{tbl:genz}). The functions $u_1$, $u_2$, $u_3$, and $u_4$ are smooth and can be approximated accurately using a polynomial of low degree. The integration error of these functions decays fastest. The fifth and sixth Genz test functions are not smooth and it is not straightforward to approximate these functions by means of a polynomial. This can also be observed in the convergence behavior of the quadrature rule, as it clearly converges slower than the other test functions.

\subsection{Two-dimensional verification load case}
\label{subsec:results2}
The two-dimensional case under consideration is a case in which the wind speed and wind direction vary, but the sea state is considered to be fixed. The distribution of the wind speed and wind direction are as described in Section~\ref{subsec:dlc12} and \ref{subsec:mmij}, so the wind speed follows the data and the wind direction is uniformly distributed between \SI{-12}{\degree} and \SI{12}{\degree}. The parameters describing the sea state are fixed at their mean values obtained from the data, which are $\Hs = \SI{1.46}{m}$, $\Tp = \SI{6.76}{s}$, and $\VWdir = \SI{-2.11}{\degree}$.

In this case, the conventional approach is to bin the wind speeds using bins of \SI{2}{m/s} and to bin the wind direction into three bins of \SI{8}{\degree}. The wind speed varies between (approximately) \SI{0}{m/s} and \SI{30}{m/s}, yielding 15 bins in this dimension (these are also depicted in Figure~\ref{subfig:binning}), resulting in a total number of 45 bins. The frequency of occurrence of each bin is determined using the number of measurements in the bin, which yields a non-zero weight for any bin. For a fair comparison, the implicit quadrature rule is constructed such that it contains 45 nodes based on the same two parameters.

For both cases, five seeds per bin (or node) are used and by means of rainflow cycle counting the weighted equivalent loads are determined. For the quadrature rule, the seeds are balanced using the procedure from Section~\ref{subsec:seedbalancing}. Seed balancing is not applied to binning, as we want to keep binning as close to the conventional approach as possible. The number of required \BLADED\ runs of the balanced seeds in conjunction with a quadrature rule decreases from 225 to 173, which is (approximately) \SI{77}{\percent} of the original cost. Some obtained loads for several components calculated using these three approaches (i.e.\ binning and quadrature rule with five seeds, and a quadrature rule with balanced seeds) are tabulated in Table~\ref{tbl:eql2}.

\begin{table}
	\caption{Equivalent loads of the two-dimensional load case, determined using various inverse S--N slopes, acting on various components of the turbine, calculated with three different methods.}
	\label{tbl:eql2}

	\sisetup{
		scientific-notation = true,
		round-mode = figures,
		round-precision = 3,
		exponent-product = \cdot,
	}

	\setlength\doublerulesep{2.0pt}

	\centering
	\resizebox{\textwidth}{!}{\begin{tabular}{lr|lll}
		\textbf{Component} & \textbf{Inv.\ S--N slope} & \textbf{Binning}        & \textbf{Quadrature rule} & \textbf{Quadrature rule} \\
		                   &                           & \small(five seeds) & \small(five seeds)  & \small(balanced seeds) \\
		\hline
		\hline
		Rotating hub & 2 & \SI{637321}{N} & \SI{636977}{N} & \SI{632210}{N} \\
		\small(In longitudinal direction) & 3 & \SI{464860}{N} & \SI{462267}{N} &\SI{459127}{N} \\
		& 5 & \SI{438793}{N} & \SI{438802}{N} & \SI{435077}{N} \\
		& 10 & \SI{492404}{N} & \SI{501829}{N} & \SI{496744}{N} \\
		& 12 & \SI{511657}{N} & \SI{523356}{N} & \SI{518813}{N} \\
		\hline
		Blade root & 2 & \SI{18498200}{Nm} & \SI{18519000}{Nm} & \SI{18469100}{Nm} \\
		\small(Flap-wise moment) & 3 & \SI{13439600}{Nm} & \SI{13410100}{Nm} & \SI{13382200}{Nm} \\
		& 5 & \SI{11554600}{Nm} & \SI{11517300}{Nm} & \SI{11475700}{Nm} \\
		& 10 & \SI{12041100}{Nm} & \SI{12070000}{Nm} & \SI{11867200}{Nm} \\
		& 12 & \SI{12492000}{Nm} & \SI{12563600}{Nm} & \SI{12264200}{Nm} \\
		\hline
		Yaw bearing & 2 & \SI{802706}{N} & \SI{801408}{N} & \SI{792923}{N} \\
		\small(In longitudinal direction) & 3 & \SI{599741}{N} & \SI{593396}{N} & \SI{587162}{N} \\
		& 5 & \SI{563348}{N} & \SI{556520}{N} & \SI{549306}{N} \\
		& 10 & \SI{640853}{N} & \SI{637199}{N} & \SI{622387}{N} \\
		& 12 & \SI{676063}{N} & \SI{670291}{N} & \SI{653069}{N} \\
		\hline
	\end{tabular}}
\end{table}

The difference between the loads determined with a quadrature rule (both determined with five seeds or balanced seeds) and those determined with binning is less than 5\%. Strictly speaking, it is not evident whether the quadrature rule or binning yields a more accurate answer in this case, as no exact values of the expected values of the loads are known. However, based on the results of Section~\ref{subsec:explicit}, we have a strong indication that the quadrature rules yield more accurate estimates. The largest variation is in the loads on the yaw bearing, which is likely due to its large sensitivity to the varying environmental conditions. The contrary is true for the blade root, which is significantly less sensitive to environmental variations. This will also be observed in the next section, where the five-dimensional case is considered.

\begin{figure}
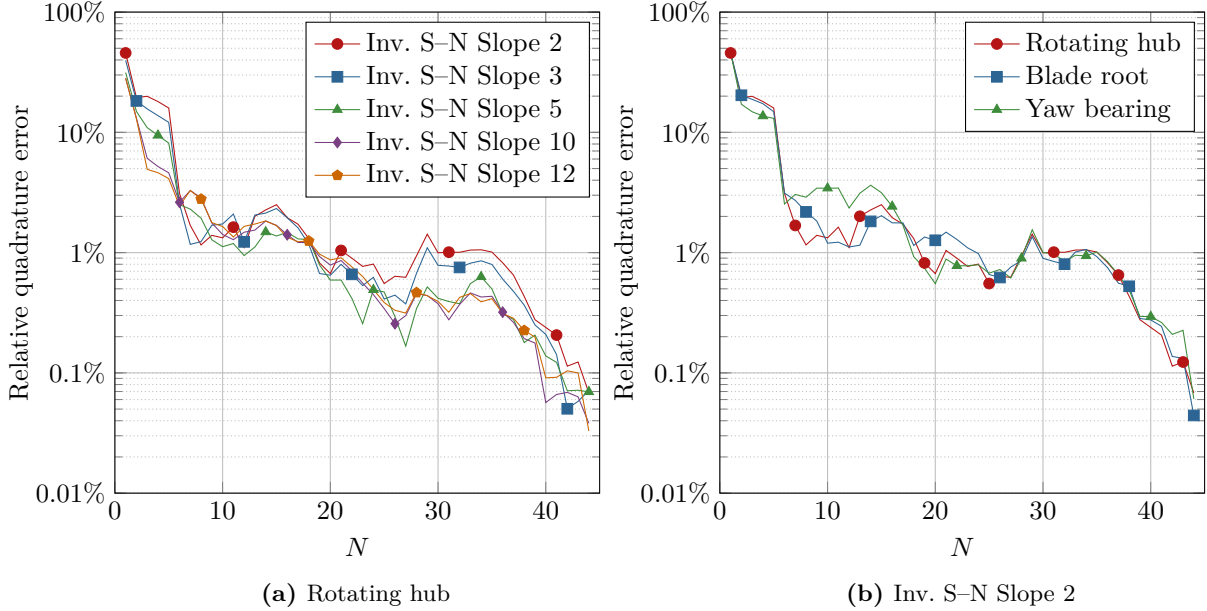

	\begin{minipage}{.5\textwidth}
		\centering
		\includepgf{\textwidth}{\textwidth}{convergence-2d-203.tikz}
		\subcaption{Rotating hub}
	\end{minipage}%
	\begin{minipage}{.5\textwidth}
		\centering
		\includepgf{\textwidth}{\textwidth}{convergence-2d-sn2.tikz}
		\subcaption{Inv.\ S--N Slope 2}
	\end{minipage}
	\caption{Relative quadrature error for various S--N slopes (left) or various components (right) considering the two-dimensional load case without seed balancing.}
	\label{fig:convergence2}
\end{figure}

One main advantage of the quadrature rule approach is that its accuracy can be assessed without any additional costly aeroelastic calculations using the procedure discussed in Section~\ref{subsubsec:implquad}. To this end, equivalent loads have been determined using $N = 1, \dots, 44$ nodes, and these are compared with the value from Table~\ref{tbl:eql2}. As discussed in Section~\ref{subsubsec:implquad} there are multiple sequences of 45 nested quadrature rules. The errors reported here are averaged over five sequences of nested quadrature rules, i.e.
\begin{equation*}
	\bar{e}_N = \frac{1}{5} \sum_{k=1}^5 \frac{| L^\mathrm{eq}_{N,k} - L^\mathrm{eq}_{45}|}{L^\mathrm{eq}_{45}}.
\end{equation*}
Here, $L^\mathrm{eq}_{N,k}$ refers to calculating the equivalent load using $N$ nodes of the $k$-th quadrature rule sequence. The error is scaled with the loads from Table~\ref{tbl:eql2}, such that loads of different order of magnitude and with different characteristics can be compared.

The obtained quadrature errors are depicted in Figure~\ref{fig:convergence2}. Notice that all reported errors decay approximately three orders of magnitude, to a point where the error is smaller than \SI{1}{\percent} of the equivalent load. The errors decay rapidly for small $N$ and the rate of decay reduces for increasing $N$. This is likely due to the fact that the error decays algebraically (as $u_5$ did in Section~\ref{subsec:explicit}), i.e.\ its decay is approximately a straight line if depicted on a log-log plot. However, significantly more \BLADED\ runs are necessary to confirm this claim numerically (e.g.\ see the number of nodes considered in the construction of Figure~\ref{fig:genz}). The left figure depicts the convergence for the equivalent loads of the rotating hub for various slopes of the S--N curve. It demonstrates that the behavior of the error is similar for different values of the S--N slope, as all lines decay rapidly. The right figure demonstrates that the behavior of the error is more or less independent from the specific component under consideration.

\subsection{Five-dimensional Design Load Case}
\label{subsec:results5}
The five-dimensional load case consists of DLC~1.2 as described in the IEC~standard~\cite{iec61400-1,iec61400-3}. The random parameters are in this case the five aforementioned parameters, consisting of wind speed, wind direction, wind/wave misalignment, significant wave height, and peak spectral period.

Applying binning to this case is infeasible, as the number of bins with non-zero frequency of occurrence equals 3\,410, which would result in approximately 15\,000 evaluations of the aeroelastic simulation code. Even after optimizing the number of seeds using seed balancing from Section~\ref{subsec:seedbalancing} the number of evaluations is still 8\,961, which remains intractable.

Instead, an implicit quadrature rule is determined consisting of 100 nodes. This number of nodes is larger than the number of nodes from the previous section, to account for the higher dimensionality of the problem. Furthermore the number of nodes is sufficient to obtain a small error (which can be assessed during the simulation using the methods discussed in Section~\ref{subsubsec:implquad}). The seeds are optimized using the method from Section~\ref{subsec:seedbalancing}, resulting in 419 individual aeroelastic simulation code evaluations. This is a major reduction compared to the number of bins, since the number of \BLADED\ evaluations is less than 5\% of that of binning. The equivalent loads acting on various components are summarized in Table~\ref{tbl:eql5}.

\begin{table}
	\caption{Equivalent loads of the five-dimensional load case, determined using various inverse S--N slopes, acting on various components of the turbine, using the full five-dimensional DLC~1.2 fatigue load case calculated with the seed-balanced quadrature rule.}
	\label{tbl:eql5}

	\sisetup{
		scientific-notation = true,
		round-mode = figures,
		round-precision = 3,
		exponent-product = \cdot,
	}

	\setlength\doublerulesep{2.0pt}

	\centering
	\begin{tabular}{r|lll}
		\textbf{Inv.\ S--N slope} & \textbf{Rotating hub} & \textbf{Blade root} & \textbf{Yaw bearing} \\
		& \small(In longitudinal direction) & \small(Flap-wise moment) & \small(In longitudinal direction) \\
		\hline
		\hline
		2 & \SI{636163}{N} & \SI{18397800}{Nm} & \SI{806393}{N} \\
		3 & \SI{466306}{N} & \SI{13379200}{Nm} & \SI{610778}{N} \\
		5 & \SI{448231}{N} & \SI{11534000}{Nm} & \SI{764622}{N} \\
		10 & \SI{549573}{N} & \SI{11991700}{Nm} & \SI{1591860}{N} \\
		12 & \SI{604697}{N} & \SI{12391400}{Nm} & \SI{1820050}{N} \\
	\end{tabular}
\end{table}

The loads of the rotating hub and blade root are for the wind turbine and offshore site under consideration relatively close to the values that were calculated in the two-dimensional load case. These loads are apparently less sensitive to environmental variations which could be due to their high position on the turbine. The loads acting on the yaw bearing are significantly larger if a random sea is considered. As the yaw bearing is the component that directly links the tower (which is connected to the sea) and the rotating hub (which is connected to the blades), its loads are sensitive to variations in both the wind and the sea. Contrary to the two-dimensional case, the sea state is random, so these variations are much larger.

\begin{figure}
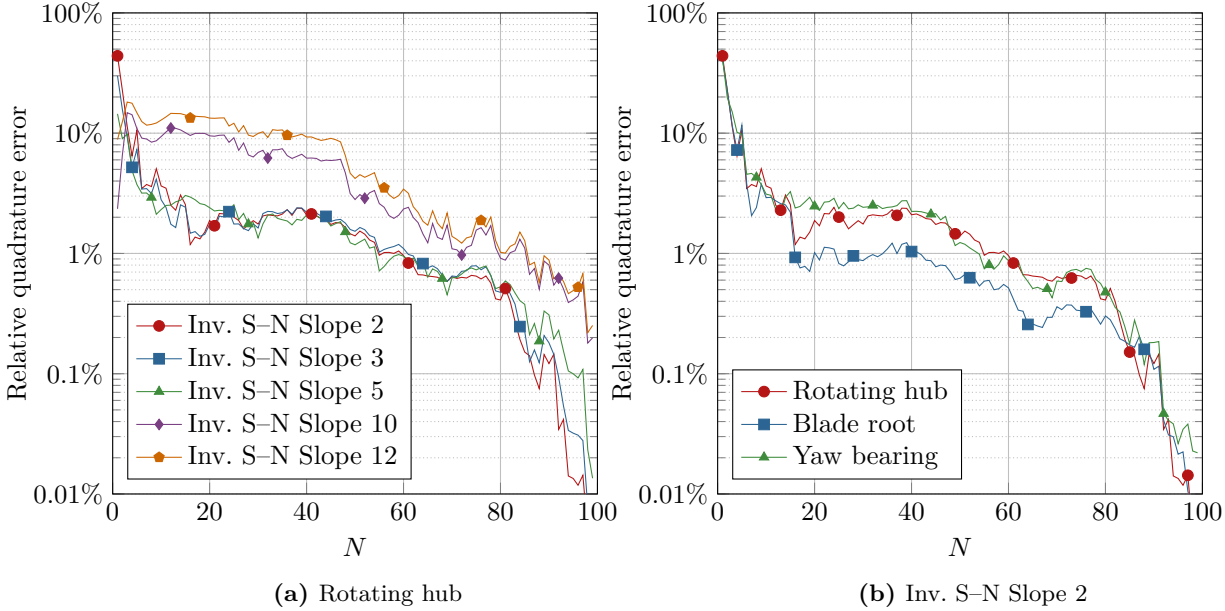

	\begin{minipage}{.5\textwidth}
		\centering
		\includepgf{\textwidth}{\textwidth}{convergence-5d-203.tikz}
		\subcaption{Rotating hub}
	\end{minipage}%
	\begin{minipage}{.5\textwidth}
		\centering
		\includepgf{\textwidth}{\textwidth}{convergence-5d-sn2.tikz}
		\subcaption{Inv.\ S--N Slope 2}
	\end{minipage}
	\caption{Relative quadrature error for various S--N slopes (left) or various components (right) considering the five-dimensional load case with seed-balancing.}
	\label{fig:convergence5}
\end{figure}

To assess the accuracy of the obtained loads, the convergence is assessed in a similar way as done in the previous section: by removal of nodes from the quadrature rule with 100 nodes, sequences of quadrature rules are constructed that are used to assess the accuracy. There is a minor subtlety in this case: the number of seeds is optimized, so the nodes of the smaller quadrature rules are evaluated using an incorrect number of seeds. We do not correct for this, so the estimated error might be slightly larger. The results are gathered in Figure~\ref{fig:convergence5}.

Compared to the two-dimensional test case, the errors of the quadrature rules are slightly larger, which is either due to the larger dimensionality of the problem or due to the unbalanced seeds used to construct these figures. Nonetheless, the errors clearly decay to a relative error smaller than 1\%. For larger slopes of the S--N curve, the error is larger, which is likely due to the higher power used in the expression that is integrated (this increases the constant of proportionality discussed in Section~\ref{subsec:seedbalancing}).  Again, the convergence behavior is possibly algebraic, though significantly more simulations are necessary to fully quantify the error. This is out of the scope of this work.

Concluding, this test case demonstrates that the proposed quadrature rule is capable of accurately approximating weighted equivalent loads in standardized form using less than 500 \BLADED\ evaluations, which is a significant reduction compared to the approximately 15\,000 \BLADED\ evaluations that binning requires.

\section{Conclusion}
\label{sec:conclusion}
A novel approach based on quadrature rules has been proposed to calculate wind turbine fatigue loads as described by Design~Load~Case 1.2, which is standardized by IEC~\cite{iec61400-1,iec61400-3}. The quadrature rule under consideration is the implicit quadrature rule, with the key properties that it has positive weights and can be constructed using measurement data. It is based on polynomial approximation, which leverages smoothness in the model to achieve high accuracy. To demonstrate the efficiency of the new approach, it has been compared to binning, the conventional approach. In both approaches, the number of seeds per bin or node can be balanced to maximize accuracy in the available computational time. The environmental parameters are based on real offshore measurements in the North Sea and the wind turbine under consideration is the NREL~5MW reference wind turbine.

Both quadrature and binning approaches have been applied to a simplified two-dimensional load case, for which it has been demonstrated that the accuracy of the quadrature rule is comparable to that of binning. Moreover, the error of the quadrature rule converges algebraically upon addition of nodes.

The main advantages of the quadrature rule, i.e.\ a significant reduction of computational time with similar accuracy, have been demonstrated by considering the full load case, which is governed by five random parameters. In this case, binning is infeasible. The accuracy of the quadrature rule has again been assessed numerically, confirming that the error of the rule decays rapidly.

Throughout this work, all results are generated based on five seeds per node. A possible improvement is to vary the number of seeds per node, for which a further study of the significance of the seed error compared to the quadrature error needs to be performed.

Overall, the results demonstrate that for fatigue load cases our proposed quadrature rule forms a highly promising alternative to binning with significantly lower computational cost and similar accuracy. Therefore, a topic for future research is to extend the numerical integration framework to other load scenarios, e.g.\ fatigue load cases with more uncertain parameters (where the benefit of using a quadrature rule is even larger) or ultimate load cases (where no rainflow cycle counting is applied).

\section*{Acknowledgments}
\label{sec:ack}
The authors acknowledge ECN part of TNO for providing the measurement data used in this work, which has been obtained as part of the Dutch Wind~Op~Zee project (\href{http://www.windopzee.net/}{\texttt{http://www.windopzee.net/}}). Lindert~Blonk and Menno~Kloosterman from DNV\,GL are acknowledged for their technical support around \BLADED\ and for providing the model of the NREL~5MW wind turbine. This research is part of the Dutch EUROS program, which is supported by NWO domain Applied and Engineering Sciences and partly funded by the Dutch Ministry of Economic Affairs.

\bibliographystyle{plainnatnourl}
\bibliography{literature.bib}

\end{document}